\documentstyle[aps,preprint]{revtex}
\begin{document}
\draft
\title{Temperature-Dependent Frequency Shifts in Collective Excitations of a 
Bose-Einstein Condensate}
\author{Hualin Shi and Wei-Mou Zheng}
\address{Institute of Theoretical Physics, Academia Sinica,\\
Beijing 100080, China}
\maketitle

\begin{abstract}
By including the contribution of the thermal cloud to the Lagrangian of the 
condensate of a Bose gas, we extend the time-dependent variational method 
at zero temperature to study temperature-dependent low collective excitation 
modes. A Gaussian trial wave function of the condensate and a static 
distribution 
density of the thermal cloud are used, and analytical expressions for 
temperature-dependent excitation frequencies obtained. Theoretical results 
are compared with measurements in the JILA and MIT experiments.

\end{abstract}


With the development of techniques to trap and cool atoms, Bose-Einstein
condensation (BEC) has been observed directly in dilute atomic vapors \cite
{and95,bra95,dav95}. The new experimental achievements have stimulated great
interest in the theoretical study of inhomogeneous Bose gases. The collective 
excitations of trapped Bose gases, which describe their dynamics and transport 
properties, have been a focus of recent experimental and theoretical 
studies. Low-lying collective excitations over a range of temperatures
\cite{jin96,mew96,jin97,sta98} and higher-lying modes\cite{and97} have 
been measured in recent experiments. 

In the regime with no detectable non-condensate 
fraction, theoretical predictions based on a mean-field description of 
weakly interacting bosons at zero temperature are in excellent agreement 
with experimental data \cite{edw96,str96,zar98}. 
At zero temperature, a Bose-Einstein condensate is described by the nonlinear
Schr\"{o}dinger equation (NLSE), i.e. the Gross-Pitaevskii equation 
\cite{gin58},
\begin{equation}
i\hbar \frac{\partial \psi }{\partial t}=-\frac{\hbar ^2}{2m}\nabla ^2\psi
+U({\bf r})\psi +\frac{4\pi \hbar ^2a}m|\psi |^2\psi , \label{eq-gp}
\end{equation}
where $m$ is the atomic mass, $U({\bf{r}})$ the trapping potential, and $a$
the $S$-wave scattering length. Characterizing the evolution of the 
macroscopic wave function $\psi $ of the condensate, this nonlinear equation 
provides a fundamental means for studying the ground state, dynamical 
behavior and collective excitations of the condensate. The validity of
the Gross-Pitaevskii equation in describing the evolution of the condensate
wave function has been verified by the quantitative comparison of 
theoretical predictions with experimental observations.

The Gross-Pitaevskii equation (\ref{eq-gp}) has been solved by means of 
a time-dependent variational technique to obtain the low energy excitations
of a trapped Bose gas for both positive and negative scattering lengths 
\cite{per96}. By taking a proper trial function of a fixed shape with some 
free time-dependent parameters, a variational technique based on Ritz's 
optimization procedure leads to a set of Newton-like ordinary differential 
equations of the second order for these parameters which characterize the 
solution to the NLSE. Avoiding expensive numerical simulations, this method 
affords analytical approximations which moreover provide a clear physical 
picture of the problem. 

For a dilute Bose gas trapped in a harmonic potential, with a Gaussian 
function taken as the natural trial wave function for low energy states, 
expressions for frequencies of the collective modes in a 3D anisotropic trap 
have been obtained \cite{per96}.
In the large particle number limit, the expressions contain the spectrum 
derived by Stringari\cite{str96} based on the Thomas-Fermi approximation, 
where the interatomic and trap interactions are dominant and the kinetic 
energy term is negligible. 

The finite-temperature excitation spectrum has also been measured in the 
experiments of Refs.\ \cite{jin97} and \cite{sta98}. The spectrum has been 
studied by means of the Popov version of the Hartree-Fock Bogoliubov 
approximation \cite{hut97,dod98} and by a two-fluid hydrodynamic approximation 
\cite{zar97,ho97}. However, the finite temperature behavior of the 
spectrum observed in experiments has not been fully understood.
In this paper, we extend the variational method at zero temperature to
finite temperature where the thermal cloud exists. At a finite 
temperature, the thermal cloud provides an additional background field 
exerting on the condensate atoms, besides the mean-field interaction 
among them. The motion of the condensate can still be described by the 
nonlinear Schr\"odinger equation with the mean-field modified to 
include the effect of the thermal cloud. The evolution equation for the 
condensate may be derived from a variational principle to minimize 
the action of the corresponding Lagrangian density ${\cal L}$.

Let us consider a sample of Bosons confined in a $d$-dimensional
harmonic potential 
\begin{equation}
U({\bf r})=\frac 12\sum_{i=1}^d m \omega_i^2 x_i^2,
\label{pot-eq}
\end{equation}
where ${\bf r}=(x_1, x_2,\ldots ,x_d)$. As shown in the recent 
experiment\cite{sta98}, we may neglect the mode coupling between the 
condensate and the thermal cloud. In this simplication, we assume the 
static thermal cloud density distribution to be
\begin{equation}
\rho _n({\bf r})=R(T) \exp [-U({\bf r})/kT],
\end{equation}
where $R(T)$ is the normalization factor. The condensate moves in the 
combined field of the external trapping potential and the mean-field potential 
due to both the condensate and the static thermal cloud. The Lagrangian 
density corresponding to the NLSE with the 
thermal cloud effect included can be written as
\begin{equation}
{\cal {L}}=\frac i2\hbar \left( \psi \frac{\partial \psi ^{*}}{\partial t}%
-\psi ^{*}\frac{\partial \psi }{\partial t}\right) +\frac{\hbar ^2}{2m}%
|\nabla \psi |^2+U({\bf{r}})|\psi |^2+\frac{2\pi a\hbar ^2}m|\psi |^4+\frac{%
4\pi \hbar ^2a}m\rho _n({\bf r})|\psi ({\bf r})|^2  \label{L-eff}
\end{equation}
where the asterisk denotes a complex conjugate. The normalization condition 
for the macroscopic wave function $\psi({\bf r})$ is
$$\int d{\bf r}|\psi |^2=N_0,$$ 
where $N_0$ is the number of the condensate atoms. The thermal cloud 
density normalizes according to
$$\int d{\bf r}\rho _n({\bf r})=N_n ,$$
where $N_n$ is the number of atoms in the thermal cloud. The last term 
in (\ref{L-eff}) comes from the interaction between the condensate and the 
thermal cloud\cite{lee57}. 

Following Ref.~\cite{per96}, we consider the Gaussian trial function
\begin{equation}
\psi ({\bf{r}},t)=A(t)\prod_i^d\exp\left\{-\frac{x_i^2}{2w_i(t)^2}+ix_i^2
\beta_i(t)\right\},
\end{equation}
where, by setting the Gaussian distribution centered at the origin, we have 
ignored the irrelevant motion of the center of the condensate. The 
variational parameters are the amplitude $A$, width $w_i $ and $\beta _i$ 
relating to the curvature. All these parameters are real time-dependent
functions, characterizing the macroscopic wave function of the 
condensate. Inserting the trial wave function into the Lagrangian density 
(\ref{L-eff}) and integrating over the coordinates, we obtain the Lagrangian
\begin{eqnarray}
L(t) &=& \langle {\cal {L}}\rangle =\int_{-\infty }^\infty {\cal {L}}
d{{\bf{r}}}\\
&=&\sum_i^d N_0 \left( \frac 14m\omega _i^2w_i(t)^2+\frac{\hbar 
^2}{4mw_i(t)^2}
+\frac{\hbar^2w_i(t)^2\beta (t)^2}{m}+\frac{1}{2} w_i(t)^2\beta
^{\prime}(t)\right)  \nonumber \\
&+&\frac{2\pi N_0^2 a\hbar ^2}{m}\prod_i^d\frac 1{\sqrt{2\pi }w_i(t)} + 
\frac{4 
\pi
N_0 N_n a \hbar^2}{m} \prod_i^d \sqrt{\frac{m \omega_i^2}{\pi k T}} \left/{
\sqrt{\frac{m \omega_i^2}{k T} w_i(t)^2 +2}}\right. ,
\label{L}
\end{eqnarray}
where we have used the equation of particle number conservation 
$$A(t)=N_0\prod_i^d\frac 1{\sqrt{\pi }w_i(t)}.$$ 
The last term in (\ref{L}) is the contribution of the interaction between 
the condensate and the thermal cloud. 

By minimizing the action corresponding to the Lagrangian with respect to 
the parameters, we can derive the ordinary equations describing the evolution
of the parameters. The widths of the condensate satisfy the following
equation: 
\begin{equation}
w''_i(t)+\omega _i^2 [1-f(T)] w_i(t)=\frac{\hbar ^2}{m^2w_i(t)^3} +\frac{4\pi
N_0 a\hbar ^2}{m^2w_i(t)}\prod_j^d\frac 1{\sqrt{2\pi }w_j(t)},  \label{w-eq}
\end{equation}
where, compared with the case of zero temperature, the thermal cloud 
contributes an extra term proportional to
$$f(T)=\frac{8 \pi^2 N_n a \hbar^2}{m^2} \left(\frac{m}{2 \pi k T}
\right)^{\frac d2 +1} \prod_{j=1}^d \omega_j.$$ 
In the derivation we have used the fact that the widths of the condensate are 
much smaller than those of the thermal cloud, i.e. $m \omega_i^2 
w_i(t)^2\ll kT $. If we regard the widths ($w_1,w_2,...w_d$) as the 
coordinates of a fictitious particle, Eq.~(\ref{w-eq}) is just Newton's 
equations of motion for the particle in the effective potential 
\begin{equation}
V_{\rm eff}=\frac 12\sum_{i=1}^d\left[ \omega _i^2 (1-f(T)) w_i^2+
\frac{\hbar^2}{m^2w_i{}^2}\right] +\frac{4\pi N_0 a\hbar ^2}{m^2}
\prod_i^d\frac 1{\sqrt{2\pi }w_i}  ,\label{veff-eq}
\end{equation}
where the $w_i^{-2}$ term corresponds to the kinetic energy of the condensate, 
the $f(T)$ term to the contribution from the interaction between the 
condensate and the thermal cloud, and the last $a$-dependent term comes 
from  the interatomic interaction of the condensate atoms. 

After finding the widths, the curvature parameters $\beta_i(t)$ can be 
obtained through the following equation derived from the variation principle
\begin{equation}
\beta _i(t)=\frac{mw'_i(t)}{2\hbar w_i(t)}.
\end{equation}
Once the widths of the condensate are known, the rest of the parameters 
can be calculated, and the evolution of the Gaussian atomic cloud of the 
condensate can then be completely characterized. The whole problem of 
solving the NLSE reduces to that of 
solving the system of ordinary differential equations (\ref{w-eq}).

As a special case, we first consider a two-dimension isotropic trap. For 
the `breath' mode with a radial symmetry, Eq.~(\ref{w-eq}) reduces to 
\begin{equation}
w''(t)+\omega ^2 (1-f_2(T)) w(t)=\frac{\hbar^2}{m^2w(t)^3}+\frac{2N_0 a\hbar 
^2}{m^2w(t)^3},
\label{w-2d}
\end{equation}
where, for $d=2$,
$$f(T)=f_2(T)={8 \pi^2 N_n a \hbar^2} (\omega/2 \pi k T)^2. $$
After introducing the new dimensionless constant $p=1+2N_0 a\hbar ^2/m^2$ 
and new dimensionless variables 
$$\tau =\omega \sqrt{1-f_2(T)}t,\qquad v=w/a_0,$$
where $a_0= [\hbar /(m\sqrt{1 - f_2(T)} \omega )]^{1/2}$, Eq.~(\ref{w-2d}) 
becomes 
\begin{equation}
v''(\tau )+v(\tau )=\frac p{v(\tau )^3}.
\label{v}
\end{equation}
This equation, regarding $v$ as the coordinate, describes the classical motion 
of a 
ficticious particle in the effective potential
$$V_{\rm eff}= v^2+p/v^2,$$
which admits an equilibrium point only when $p>0$. In this case 
Eq.~(\ref{v}) has the analytical solution \cite{kag97} 
\begin{equation}
v^2(\tau ) = \frac 12\left[v_0^2+\frac p{v_0^2}\pm \left(v_0^2-\frac 
p{v_0^2}\right)\cos (2\tau )\right].
\end{equation}
The oscillation frequency is $2\omega \sqrt{1-f_2(T)}$. For a two-dimensional
harmonic trap, $N_n \sim T^{5/2}$ \cite{gh50,bag87,shi97b}, so the temperature 
dependency of the frequency is 
$$\omega(T) \sim \sqrt{1-C T^{1/2}},$$ 
where $C$ is a constant independent of $T$. At zero temperature the frequency 
reduces to $2 \omega$, as is expected. For a positive scattering length, the 
frequency decreases with increasing temperature. On the contrary, for a 
negative scattering length, the frequency increases with temperature. At a 
negative scattering length $p$ becomes negative when $N_0>m^2/(2 |a| \hbar^2)$. 
This gives a critical value of the atom number beyond which no condensate 
would occur, nor is there any oscillatory solution. This critical value is 
independent of the temperature. 

In the case of a three-dimensional cylindrically symmetric harmonic trap 
\begin{equation}
U({\bf r})=\frac 12 m \omega^2(x^2+y^2+\lambda^2z^2),
%\label{pot-eq}
\end{equation}
the equations for the widths, in terms of the dimensionless constant 
$P_0=\sqrt{2/\pi }N_0(T)a/a_0$ and new dimensionless variables 
$\tau =\omega t$, $\nu_i=w_i/a_0$ with $i=x,y,z$ and $a_0=\sqrt{\hbar 
/m\omega }$, are 
\begin{eqnarray}
\frac{d^2\nu _x}{d\tau ^2}+\left[ 1-f_3(T)\right] \nu _x &=&\frac 1{\nu _x^3}%
+\frac{P_0}{\nu _x^2\nu _y\nu _z}  \label{m3x-eq} ,\\
\frac{d^2\nu _y}{d\tau ^2}+\left[ 1-f_3(T)\right] \nu _y &=&\frac 1{\nu _y^3}%
+\frac{P_0}{\nu _y^2\nu _x\nu _z}  \label{m3y-eq} ,\\
\frac{d^2\nu _z}{d\tau ^2}+\left[ \lambda ^2-f_3(T)\right] \nu _z &=&\frac 1{%
\nu _z^3}+\frac{P_0}{\nu _z^2\nu _x\nu _y} . \label{m3z-eq}
\end{eqnarray}
Here, introducing the notation $\eta =T/T_c$, $\bar{\omega}=\lambda^{1/3}
\omega$ and $\bar{a}_0=\sqrt{\hbar/m \bar{\omega}}$, we have denoted for $d=3$
$$f_3(T)=f(T)= \eta^{\frac 52} N_n(T)\frac{a}{\bar{a}_0}\left(\frac 2 
\pi\right) 
^{\frac 12}\left[\frac {\varsigma (3)}{N_0(T_c)}\right]^{\frac 56}.$$
The $f_3(T)$ terms come from the interaction between the condensate and the
thermal cloud. The $\nu_i^{-3}$ terms correspond to the kinetic energy 
contribution which is neglected in the Thomas-Fermi approximation but plays 
a role in stabilizing the condensate. As in the case of zero temperature, for a 
positive scattering length $a>0$, the effective 
potential has a stable equilibrium point, while for a negative
scattering length $a<0$ there may be no equilibrium points, depending on 
$N_0$ in comparison with a critical value $N_c$. For an attractive 
interatomic interaction, $N_c$ sets an upper limit of the atom number, 
i.e. a collapse occurs when $N_0>N_c$. When $N_0<N_c$, a metastable 
equilibrium point exists as one of the two equilibrium points, supporting 
the condensate. In contrast to the two-dimensional case, for a 
three-dimensional trap, the critical atom number increases with
temperature. This is consistent with what is known in the case of an isotropic
trap \cite{shi97a}.

Following Ref.~\cite{per96}, we may study the collective excitation modes of 
the condensate by expanding $\nu_i$ of Eqs.~(\ref{m3x-eq})-(\ref{m3z-eq})
around the stable equilibrium point ($\nu _0,\nu _0,\nu _{0z}$), which 
satisfies
\begin{eqnarray}
\left[ 1-f_3(T)\right] \nu _0 &=&\frac 1{\nu _0^3}+\frac{P_0}{\nu _0^3\nu 
_{0z}%
} ,\\
\left[ \lambda ^2-f_3(T)\right] \nu _{0z} &=&\frac 1{\nu _{0z}^3}+\frac{P_0}{%
\nu _0^2\nu _{0z}^2}.
\end{eqnarray}
A linear analysis then leads to the following expressions
for the low excitation frequencies
\begin{eqnarray}
\omega (|m|=2) &=&2\omega \sqrt{1-f_3(T)-2P_{4,1}}  \label{f2-eq} ,\\
\omega (|m|=0) &=&\sqrt{2}\omega \left[ \left(1-f_3(T)\right)
\left(1+\lambda^2(T)\right)-P_{2,3} \phantom{\sqrt{P_2^2}}\right. \nonumber\\
& &\left. \pm \sqrt{\left[\left(1-f_3(T)\right) 
\left(1-\lambda^2(T)\right) +P_{2,3}\right]^2+8P_{3,2}^2}\right] ^{\frac 12},  
\label{f0-eq}
\end{eqnarray}
where $P_{i,j}=P_0/(4\nu _0^i\nu _{0z}^j)$ and $\lambda^2 (T) =(\lambda
^2-f_3(T))/(1-f_3(T))$, and we have labeled the modes by the azimuthal 
angular quantum numbers $m$. By noticing that $f_3(T)=0$ at zero 
temperature, expressions (\ref{f2-eq}) and (\ref{f0-eq}) reproduce the 
results obtained in Ref.\cite{per96}. 

For small $P_0$, expressions (\ref{f2-eq}) and (\ref{f0-eq}) are close to the 
bare trap results. For large values of $P_0$ they correspond to the case 
of large interatomic interactions in comparison with the trap excitation 
energies. They are good for either the repulsive or attractive interatomic 
interactions, i.e. for both the positive and negative scattering lengths. 
In the large atom number limit, where the kinetic 
energy term is negligible, expressions (\ref{f2-eq}) and (\ref{f0-eq}) can 
be further simplified as
\begin{eqnarray}
\omega (|m|=2) &=&\omega \sqrt{2(1-f_3(T))},  \label{m2tf-eq} \\
\omega (|m|=0) &=&\omega \sqrt{\frac{1-f_3(T)}2}\left[ 4+3\lambda^2 (T)\pm 
\sqrt{16+9\lambda^4(T)-16\lambda^2 (T)}\right] ,  \label{m0tf-eq}
\end{eqnarray}
which, with  $(1-f_3(t))\omega $ replaced by $\omega $ and $(\lambda
^2-f_3(t))/(1-f_3(T))$ by $\lambda ^2$ at zero temperature, correspond to 
the Thomas-Fermi approximation. Near the critical temperature $N_0$ is 
small, so the interatomic interaction of the condensate may not be dominant 
over the kinetic energy. For attractive interacting Bose gases of a negative 
scattering length, no condensate exists without the contribution of the 
kinetic energy being included. In these cases, instead of expressions 
(\ref{m2tf-eq}) and (\ref{m0tf-eq}), expressions (\ref{f2-eq}) and 
(\ref{f0-eq}) should be used.

In the MIT experiment \cite{sta98}, the axial trapping frequency $\nu_z 
=16.93$Hz and the radial frequency $\nu _r=230$Hz. Without considering the 
interatomic interaction, the critical temperature $kT_c=\hbar \omega (N\lambda 
/1.202)^{1/3}$ \cite{gh50,bag87,shi97b} for $N=80\times 10^6$ is $1.87\mu$K, 
which is close to the experimental value $T_c=1.7\mu$K. The 
difference may come from the neglected atomic mutual interaction 
\cite{shi97b,gio96}. According to the local density approximation the 
chemical potential is given by \cite{shi97c} 
\begin{equation}
\mu =\frac{8\pi a\rho _0\hbar ^2}m+\frac{\hbar \bar{\omega}}2\left\{ 
\frac{15aN%
}{\bar{a_0}}\left[ 1-\left( \frac T{T_c}\right) ^3\right] \right\} ^{\frac 
25},
\end{equation}
which, for very low temperatures, reduces to 
$$\mu =\frac{\hbar \bar{\omega}}{2} \left[\frac{15 a 
N}{\bar{a_0}}\right]^{\frac 25}.$$ 
Noticing that $a=2.75$nm in the MIT experiment, we have the chemical potential 
$\mu /k_{\rm B} \approx 348$nk at $N_0=15\times 10^6$ in agreement with the 
experimental value 380nK. This verifies 
the validity of the local density approximation.

Using the experimental parameters we have calculated the oscillation 
frequencies. For the experiment on Rb vapor at JILA, the particle number is 
not too large, so expressions (\ref{f2-eq}) and (\ref{f0-eq}) are used. 
The temperature dependent excitation spectrum from the theoretical calculation 
together with the experimental measurements is shown in Fig.~1. It is seen 
that at low temperatures the frequencies decrease with increasing temperature. 
A simple theory explains the experimental data at low temperature when the 
condensate occupies a significant proportion. In the MIT experiment 
for mode $m=0$, the atom number is rather 
large, so we use expressions (\ref{m2tf-eq}) and (\ref{m0tf-eq}). The 
comparison of the theory with experiment is shown in 
Fig.~2. The theoretical results are in good agreement with experimental 
observation.

In summary, by including a mean field acting on the condensate from the 
static thermal cloud in the Lagrangian density, we have studied the 
collective excitation modes of BEC 
at finite temperature by means of the time-dependent variational method, and 
obtained analytic expressions for the temperature-dependent frequency shift 
of low excitation modes. We have compared the theory with experimental 
observations. The theoretical results are in agreement with the MIT experiment 
very well.  

\acknowledgments{This work was supported in part by the National Natural
Science Foundation of China.}

\begin{figure}
\caption{Temperature-dependent excitation spectrum of the JILA experiment 
for collective modes $m=0$ (triangle) and $m=2$ (circle), compared with 
theoretical curves (dotted for $m=0$ and solid for $m=2$). Frequencies are 
normalized with respect to the radial trap frequency, and temperature to 
the critical temperature for a harmonically confined ideal gas}  
\label{fig1}
\end{figure}

\begin{figure}
\caption{Temperature-dependent collective excitation frequencies of the MIT 
experiment for mode $m=0$ (circle), compared with the theoretical curve. In the 
experiment, temperature was varied by adjusting the rf frequency of a magnetic 
trap, so here $\Delta \nu_{\rm rf}$ is a measure of temperature.}   
\label{fig2}
\end{figure}

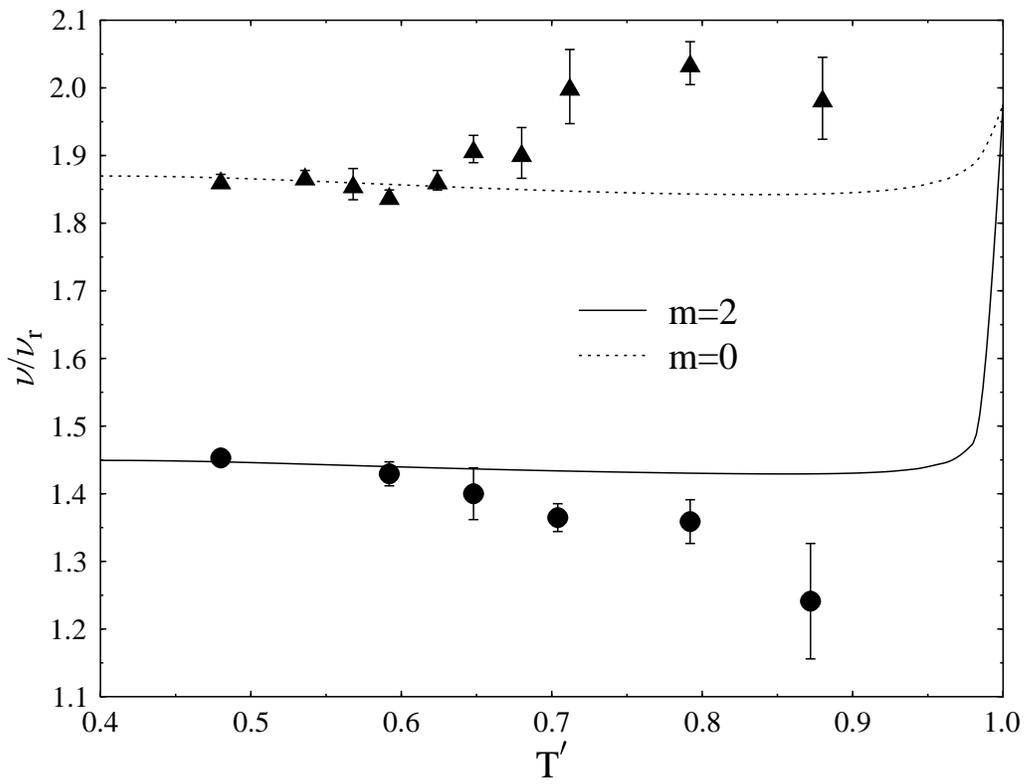
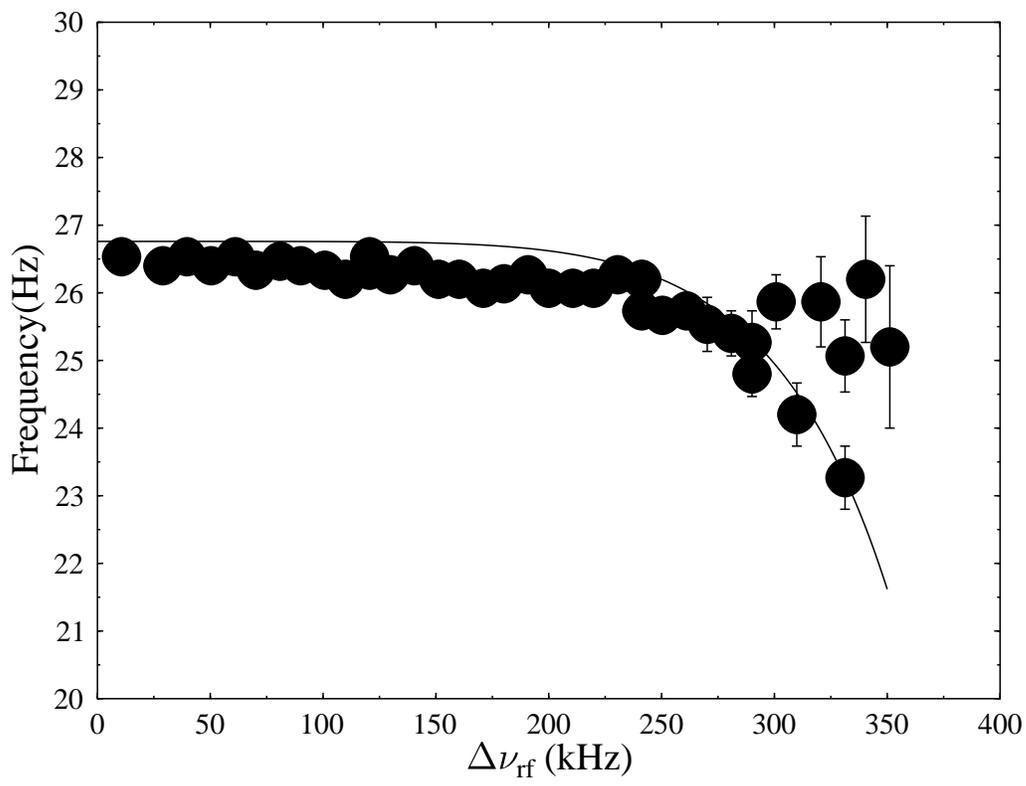
\end{document}